\documentclass[a4paper]{article}
\usepackage{latexsym}
\begin{document}

\newtheorem{theorem}{Theorem}
\newtheorem{proposition}{Proposition}
\newtheorem{remark}{Remark}
\newtheorem{corollary}{Corollary}
\newtheorem{lemma}{Lemma}
\newtheorem{observation}{Observation}

\newcommand{\qed}{\hfill$\Box$\medskip}

\title{A Non-Oblivious Reduction of\\Counting Ones to Multiplication} 
\author{Holger Petersen\\ 
Reinsburgstr. 75\\
70197 Stuttgart\\
Germany} 

\maketitle

\begin{abstract}
An algorithm counting the number of ones in a binary word is presented running
in time $O(\log\log b)$ where $b$ is the number of ones. The operations available include
bit-wise logical operations and multiplication.
\end{abstract}

\section{Introduction}  
The operation of counting the number of ones in a binary word consisting of $n$ bits 
has received considerable attention in quite different fields such as Cryp\-to\-gra\-phy \cite{LM01}
and Chess Programming \cite{Chess}, where
the operation is used to evaluate the legal moves a player has in a given position.
Counting ones is also known under the names sideways addition \cite{Knuth09}, 
bit count \cite{KR78}, or population count \cite{Chess}. 

An early reference describing a non-trivial method for counting ones 
is the article by Wegner \cite{Wegner60}:
Instead of looping through all $n$ bits of a machine word, the right-most one 
of an operand $x > 0$ is repeatedly deleted by the operation $x\mbox{ and } (x-1)$, where ``$\mbox{and}$''
denotes a bit-wise operation. In this way complexity $O(\nu x)$ is achieved, where $\nu x$ is the 
number of ones in the input $x$. This technique is also suggested in Exercise~2-9 of \cite{KR78}.

By forming growing blocks of bits, complexity $O(\log n)$ can be achieved with the help of 
constant time shift operations \cite{Chess}.
Under unit cost measure for multiplication or division, 
algorithms of asymptotical time complexity $O(\log\log n)$
are the Gillies-Miller method \cite{WWG57,Knuth09} and Item~169
of \cite{HAKMEM}. 

In contrast to Wegner's approach, the asymptotically more efficient
solutions are oblivious in the sense that their complexity is independent of the input value.
A sparse input (containing few ones) 
is not processed more efficiently than an input with many ones.
If, e.g., the input is known to contain at most a constant number of ones, then Wegner's
method has time complexity $O(1)$.

In this note we show that the Gillies-Miller method can be modified to work in a
non-oblivious way.

\section{Result}
\begin{theorem}
Counting ones can be done in $O(\log\log b)$ steps under 
unit cost measure for logical and arithmetical operations including multiplication,
where $b = \nu x$ is the number of ones in the input $x$.
\end{theorem}
{\bf Proof:}
We describe an algorithm that uses several families of constants (''magic masks'' in the sense of \cite{Knuth09}) that potentially extend infinitely towards higher order bits. 
In concrete implementations these constants can be truncated to the 
word length of the processor architecture and only a finite number of values is required.

The first family $\mbox{\tt m}$ of masks selects blocks of  bits depending on parameter $k$:
$$\mbox{\tt m[k]} = \cdots \underbrace{11\cdots 11}_{2^k} \underbrace{00\cdots 00}_{2^k} \underbrace{11\cdots 11}_{2^k}$$

The second family $\mbox{\tt h}$ selects the most significant bits from each block:
$$\mbox{\tt h[k]} = \cdots \underbrace{10\cdots 00}_{2^k} \underbrace{10\cdots 00}_{2^k} \underbrace{10\cdots 00}_{2^k}$$
The masks $\mbox{\tt h}$ are also used in a modified form as multipliers for adding up blocks of  bits
of the current value of  $\mbox{\tt x}$.

Finally we make use of the table $\mbox{\tt e[k]}$ with $\mbox{\tt e[k]} = 2^k$.

Each iteration of the while-loop in code that follows starts with $x$ consisting of a sequence of blocks of length $\ell = 2^k$, 
where each block contains the number of ones of the input in the corresponding bit positions. Variable $p$ 
holds the product of  $x$ and  $2h[k+1] + 1$. The test concerning $p \mbox{ and } h[k]$ determines if
$\ell-1$ bits suffice to hold the count of all ones in the input. In fact, the test could be a little less strict 
with respect to the most significant bit, which may be a 1 in the block containing the count of all blocks. 

\begin{verbatim}
function bitcount(x: integer): integer;
var k, p: integer;
begin
  k := 0;
  p := -x;
  while (p and h[k]) <> 0 do
  begin
    x := (x and m[k]) + ((x div e[k+1]) and m[k]);
    p := x * (2*h[k+1] + 1);
    k := k+1
  end;
  bitcount := (p div e[(n div 2) - e[k]]) and (e[e[k]]-1)
end;
\end{verbatim}

We now argue that the above algorithm determines whether an 
overflow to the most significant bit of the blocks being added up
occurs.
Let $\ell = 2^k \ge 1$ be a block-length and consider the current
$x$ as a sequence of blocks $x_{n/\ell}\cdots x_{1}$ 
of length $\ell$. We claim that if 
$\sum_{i=1}^{n/\ell} \ge 2^{\ell}$ then there is a $j \le n/\ell$
such that $2^{\ell-1} \le \sum_{i=1}^{j} \le 2^{\ell}-1$.
If $x_1 \ge 2^{\ell-1}$ then we can take $j = 1$ since 
$\ell \le 2^{\ell-1} \le  2^{\ell}-1$ for $\ell \ge 1$. Otherwise there is a maximum
$s$ such that $\sum_{i=1}^{s} \le 2^{\ell-1}-1$. 
Then $\sum_{i=1}^{s+1} \le 2^{\ell-1}-1 + 2^{\ell-1} = 2^{\ell}-1$
and we can set $j = s+1$. Now $\sum_{i=1}^{j}$ has bit $\ell$
set such that ``$p \mbox{ and } h[k]$'' will not be 0. 

When the loop terminates, the Gillies-Miller method will have produced 
the sum of all blocks in the ``middle''  block of $p$ \cite{WWG57}. This block
is extracted by the last assignment.

Since the loop terminates with the smallest $k$ such that $2^{\ell-1} = 2^{2^k-1} > b$ 
we have $k = O(\log\log b)$.
\qed


\bibliographystyle{abbrv}

\end{document}